\documentclass[a4paper,fleqn]{article}
 \usepackage{amsmath,amsfonts,cite}
\usepackage[nodayofweek]{datetime}

\def\bea{\begin{eqnarray}}
\def\eea{\end{eqnarray}}

\newtheorem{theorem}{Theorem}[section]

\begin{document}

\noindent{\bf \Large Lie point symmetries and ODEs passing the Painlev\'e test } \\

\noindent {\bf   D. Levi$^*$, D. Sekera$^{\dag}$, P. Winternitz$^{\ddag}$ }

\noindent{\bf  $^*$ 
Dipartimento di Matematica e Fisica, Roma Tre University and   INFN,  Sezione  Roma Tre, Via della Vasca Navale 84, 00146 Roma, Italy, $\qquad$
  e-mail: levi\@roma3.infn.it\\
  $^{\dag}$ Institute of Physics, Laboratory for Particle Physics and Cosmology,
\'Ecole Polytechnique F\'ed\'erale de Lausanne, CH-1015 Lausanne, Switzerland, $\qquad$
e-mail : david.sekera@epfl.ch\\
  $^{\ddag}$D\'epartement de math\'ematiques et de statistique and  Centre de recherches math\'ematiques,
Universit\'e de Montr\'eal, C.P. 6128, succ. Centre-ville, Montr\'eal (QC) H3C 3J7,
Canada, \\ e-mail:
wintern@crm.umontreal.ca                    

}

\begin{center}
{\bf \shortdate
\today}
\end{center}
\begin{abstract}
 The Lie point symmetries of ordinary differential equations (ODEs) that are
candidates for having the Painlev\'e property are explored for ODEs of
order $n =2, \dots ,5$.  Among the 6 ODEs identifying  the Painlev\'e
transcendents only $P_{III}$, $P_V$ and $P_{VI}$ have nontrivial symmetry algebras and
that only for very special values of the parameters. In those cases the
transcendents can be expressed in terms of simpler functions, i.e. elementary functions, solutions of linear equations, elliptic functions or Painlev\'e transcendents occurring at lower order.
For higher order or higher degree ODEs that pass the Painlev\'e test only
very partial classifications have been published. We consider many
examples that exist in the literature and show how their symmetry groups
help to identify those that may define genuinely new transcendents.
\end{abstract}
\section{Introduction}
Two systematic methods exist for solving nonlinear ordinary differential equations 
(ODEs) or systems of ODEs. One is symmetry analysis, based on Lie group theory. The other is based on the study of the singularity structure of the solutions of the considered ODEs, namely Painlev\'e analysis.

The purpose of this article is to study the interplay between the two methods when both are applicable. Thus we shall consider one nonlinear ODE of the form
\bea \label{1.1}
y^{(n)}=R(x, y, y', \cdots, y^{(n-1)}),
\eea
where $R$ is rational in $y, y', \cdots, y^{(n-1)}$ with coefficients that are analytic in $x$.  We shall assume that \eqref{1.1} passes the Painlev\'e test in its original form \cite{1,2,2a,3,c,c1,gromak}, that is that its general solution can be written in the form
\bea \label{1.2}
y(x)= \sum_{n=0}^{\infty} \, a_n \, (x - x_0)^{n+\alpha}, \qquad a_o \ne 0, \quad x_0 \in \mathbb C,
\eea
were $\alpha$ is an integer (usually a negative one) and $x_0$ is an arbitrary point in the complex plane. 
The constants $a_n$ are to be determined from a recursion relation of the form
\bea \label{1.3}
P(k) a_k = \phi(a_{k-1}, \cdots, a_0, x_0),
\eea
where $P(k)$ is a polynomial in $k$ that has $n-1$ non-negative integer roots, called resonances. The values of $a_k$ for $k$ at a resonance is arbitrary, but a resonance condition
\bea \label{1.4}
 \phi(a_k, a_{k-1}, \cdots, a_0, x_0, \alpha)=0
\eea
must be satisfied for each resonance value of $k$, identically in $x_0$ and in the values of $a_{\tilde k}$ where $\tilde k < k$ are all previous resonance values.
If $n-1$ such resonances exist then the solution \eqref{1.2} depends on $n$ arbitrary constants $(x_0, a_k, a_{k_1}, a_{k_2}, \cdots, a_{k_{n-1}})$ and may be the general solution. If \eqref{1.1} is of the {\it polynomial type}, i.e. $R$ is a polynomial in $y$ and its derivatives, the Painlev\'e test is complete. If $R=\frac P Q$ is the ratio of two polynomials $P$ and $Q$, then further expansions are necessary to investigate values of $x_0$ where the polynomial in the denominator, $Q(y, y', \cdots, y^{(n-1)}, x)$ vanishes.

The test provides necessary conditions for having the Painlev\'e property.  To make it sufficient one would have to prove that the radius of convergence of the series \eqref{1.2} is positive and that arbitrary initial conditions can be satisfied.

Let us now assume that \eqref{1.1} passes the Painlev\'e test. The question is what further can be achieved by performing a symmetry analysis.

We will consider examples of ODEs of order up to 5 that pass the Painlev\'e test  and show how a symmetry analysis can help to identify those for which we can reduce their order or to find particular or general solutions. In cases when the ODEs involve parameters, symmetry analysis can identify parameter values for which the order of the ODEs can be reduced.

Part of our motivation is that in the study of superintegrable systems, i.e. systems with more integrals of motion than degrees of freedom {\it exotic potentials} keep occurring \cite{4,5,6,7,8,9,10}. These are potentials that satisfy nonlinear ODEs of order $N$, where $N$ depends on the order of the integrals of motion as quantum mechanical operators. So far for integrals of order 3, 4, 5 it turns out that all exotic potentials that allow the representation of variables in $E_2$ in Cartesian or polar coordinates satisfy ODEs that pass the Painlev\'e test.

In Section 2 we shall look at the symmetry properties of the 50 classes
of second order first degree ODEs with the Painlev\'e property obtained by
Painlev\'e and Gambier. Examples of higher order ODEs are treated in
Section 3. Throughout the text the equations and their solutions can be real or
complex. Correspondingly their symmetry algebras will be considered over
the field $\mathbb F= \mathbb R$ or $\mathbb F=\mathbb C$.
\section{Second order ODEs with the Painlev\'e property and their Lie point symmetries}
First of all some comments on Lie point symmetries of ODEs are in order. The symmetry algebra of a given ODE will be realized by vector fields of the form
\bea \label{2.1}
\hat X= \xi(x,y) \partial_x + \phi(x,y) \partial_y
\eea
where $\xi(x,y)$ and $\phi(x,y)$ are smooth functions on some open set $\{ x,y \} \in \mathbb C^2$. For a given ODE
\bea \label{2.2}
F(x, y, y', \cdots, y^{(n)})=0,
\eea
one obtains the functions $\xi(x,y)$ and $\phi(x,y)$ by solving the {\it determining equations} obtained using a standard algorithm \cite{Lie,Olver,Bluman}.  For $n\ge 2$ the determining equations are an overdetermined system of linear partial differential equations (PDEs).

For second order ODEs the following results (due to Sophus Lie) are useful for our purposes.
\begin{theorem}(Lie) A second order ODE
\bea \label{2.3}
F(x,y, y', y'')=0, \quad \frac{\partial F}{\partial y''} \ne 0,
\eea
can have a symmetry algebra of dimension $dim L=N$, with $N=0,1,2,3,$ or $8$. For $N=1$ equation \eqref{2.3} can be reduced to a 1st order ODE. For $N=2$ equation  \eqref{2.3} can be solved by quadratures. For $N=3$ the quadratures can be calculated explicitly. For $N=8$ the algebra is locally isomorphic to $sl(3)$ and \eqref{2.3} can be transformed into the equation $\ddot u(t)=0$ (by a point transformation $(x,y) \rightarrow (t,u)$).
\end{theorem}
Comments
\begin{enumerate}
\item For $N=1, 2, 3$ the obtained solutions may be implicit.
\item Every second order ODE linearizable by a point transformation has $dim L = 8$
\end{enumerate}

\noindent For $n \ge 3$ we have the following results:
\begin{enumerate}
\item In \eqref{2.2} the maximal dimension of its symmetry algebra is $N=n+4$. This is achieved for the equation 
\bea \label{2.4}
y^{(n)}=0
\eea
and the corresponding vector fields are $$\{ \partial_x,\,x\partial_x,\,x^2\partial_x + (n-1)xy \partial_y, \, y\partial_y, \, \mbox{and}\, x^k \partial_y, \;k=0,\cdots, n-1 \}.$$
It is however not true that any linear ODE of order $n$ has $dim L=n+4$ and only a subclass of linear ODEs can be transformed into \eqref{2.4}, namely precisely those with $dimL =N+4$.  Other possible
dimensions for symmetry algebras of $n^{th}$ order linear ODEs are  $ n+1$ and $n+2$.
\item If we have  $dim L=1$ or $2$, the order of \eqref{2.2} can be reduced by $1$ or $2$, respectively. If we have $dim L=N \ge 3$, the order of \eqref{2.2} can be reduced by $s \le N$ where $s$ is the dimension of a  maximal solvable subalgebra of $L$.
\end{enumerate}

Painlev\' e \cite{11} and Gambier \cite{12} classified all equations of the form \eqref{1.1} for $n=2$ that have the Painlev\' e property (no movable singularities other than poles) into $50$  conjugacy  classes under the (local) Lie group of (local) point transformations of $(x,y)$ that preserve the Painlev\' e property.  These are arbitrary locally invertible transformations of $x$ and fractional linear transformations of $y$ $(x,y) \rightarrow (\tilde x, \tilde y)$ with
\bea \label{2.5}
\tilde x = \phi(x), \quad \tilde y = \frac{\alpha(x) y + \beta(x)}{\gamma(x) y + \delta(x)}, \quad \alpha \delta - \beta \gamma \ne 0,
\eea
where $\phi, \, \alpha,\,\beta,\,\gamma,\,\delta$ are arbitrary functions, smooth in some interval $x \in I$, such that the inverse transformation exists.

Among these $50$ classes, listed in many books (see e.g. \cite{14,15}), $6$ define the famous Painlev\' e trascendents. All $50$ equations, in their standard form, can be written as
\bea \label{2.6}
y''=\mathcal L(x,y) (y')^2+\mathcal M(x,y) y'+\mathcal N(x,y),
\eea
where $\mathcal L$ is rational in $y$ and has between 1 and 4 poles
in $y$. One pole at $y=y_0$ is shifted to $y=\infty$. Then $\mathcal L$ takes one of the forms \bea \label{2.6a} \mathcal L=\Big \{ 0, \frac a y, \frac b y + \frac c {y-1}, \frac 1 2 \Big [ \frac 1 y + \frac 1 {y-1} + \frac 1 {y-\lambda} \Big ] \Big \}\eea
where $\lambda$ is arbitrary, but the residues $a$ and $b$  have specific values  (like $a=1$, or $a=\frac {m-1} m $ with $m \ge 2$ and integer).  The list of $50$ equations is standarly ordered  into $8$ classes by increasing order of the number of poles in $\mathcal L$ as in \eqref{2.6a}. 

We have calculated the Lie point symmetries for all $50$ classes in the list and for our purposes a different order emerges (we follow the standard list given e.g. in Ince \cite{15} and our notation for the classes of equations in Ince's list is $I_1, I_2, \cdots, I_{50}$.
\begin{enumerate}
\item {\bf Autonomous equations}.  For these the dimension $N$ of the symmetry algebra $L$ satisfies $N=1,2,3$ or $8$. Indeed the Lie algebra always contains the vector field $\hat X_1=\partial_x$ ($dim L=1$). If $dim L=8$ the algebra is isomorphic to $sl(3,\mathbb C)$ and the equation can be transformed into $\ddot u=0$. The ODEs in this class and the values of $N=dim L$ are: $I_1, \; (N=8)$; $I_2, \; (N=2)$; $I_3,\; (N=1)$; $I_7, \; (N=2)$; $I_8, \; (N=1)$; $I_{11}, \; (N=8)$; $I_{12},\; (N=1)$; $I_{17}, \; (N=8)$; $I_{18}, \; (N=2)$; $I_{19}, \; (N=1)$; $I_{21}, \; (N=2)$; $I_{22}, \; (N=3)$; $I_{23}, \; (N=1)$; $I_{29}, \; (N=2)$; $I_{30}, \; (N=1)$; $I_{32}, \; (N=3)$; $I_{33}, \; (N=1)$; $I_{37}, \; (N=8)$; $I_{38}, \; (N=1)$; $I_{41}, \; (N=8)$; $I_{43}, \; (N=8)$; $I_{44}, \; (N=1)$; $I_{49}, \; (N=1)$.
\item {\bf Equation with explicit (polynomial) dependence on the independent variable $x$}. $I_4 \sim P_I$, $I_9 \sim P_{II}$, $I_{13} \sim P_{III}$, $I_{20}$ equivalent to special case of $P_{II}$, $I_{31}\sim P_{IV}$, $I_{34}$ reducible to $P_{II}$, $I_{39} \sim P_{V}$, $I_{50}\sim P_{VI}$. We see that all the irreducible Painlev\' e transcendents appear in this class. All of them, except $P_I$, depend on at least one parameter. For generic values of the parameters we have $N=0$. For specific values of the parameters there are some exceptions (see below).
\item {\bf Equations involving arbitrary functions of $x$}. These are equations, $I_5,\,I_6,\,I_{15},\,I_{16}$ and $I_{24}$ with one arbitrary function $q(x)$ and $I_{14},\,I_{25},\, I_{27}$ and $I_{40}$ with two arbitrary functions each. Here Lie group theory is of little help, but Painlev\' e and Gambier integrated these equations by finding a first integral or presenting a non-point transformation $y=g(x,u, \dot u)$ that ultimately reduces the given equation to a linear equation, or one for elliptic functions.
\item {\bf Equations involving functions that satisfy specific differential equations}. Let us run through all cases.
\bea \label {2.7}
{\bf  I_{10} } \qquad y''= -y y' +y^2 -12 q y +12 q'
\eea
\begin{enumerate}
\item $q=0,\; N=2$
\item $q=q_0\ne 0, \; (q_0=\mbox{const.}),\; N=1$
\item $q''=6q^2, \; q \ne 0, \; N=1$
\item $q''=6 q^2 +1, \; q \ne q_0, \; N=1$
\item $q''=6 q^2 + x, \; N=0$
\end{enumerate}
In cases (a) and (b) $I_{10}$ has constant coefficients and can be integrated by symmetry. In cases (c) and (d) the function $q$ is an elliptic function and $I_{10}$ can be reduced to a first order ODE (and to elliptic functions). In case (e) the function $q$ is itself a $P_I$ transcendent  and $I_{10}$ is solved in terms of $P_I$.
\bea \label{2.8}
{\bf I_{26} } \qquad y'' = \frac 3 {4y} (y')^2 + \frac{6q'} y y' + 3 y^2 +12 q y -12 q'' - \frac{36 (q')^2} y
\eea
 \begin{enumerate}
\item $q''=6q^2, \; q \ne 0, \; N=1$
\item $q''=6 q^2 +1/2,  \; N=1$
\item $q''=6 q^2 + x, \; N=0$
\end{enumerate}
Here again, in cases (a) and (b) $q$ is an elliptic function and the order of the equation can be decreased by one. In case (c) $q$ is a $P_I$ transcendent and so is the solution of $I_{26}$.

The remaining cases $I_{28}, \, I_{35},\, I_{36},\, I_{42},\, I_{45},\, I_{46},\,I_{47},\,I_{48}$ are quite similar. The ODEs involve one or more functions of $x$ that are all ultimately expressed in terms of one function. This function either satisfies  an equation for elliptic functions, or for one of the irreducible Painlev\' e transcendents, namely $P_I$, $P_{II}$ or $P_{IV}$. In the first case a nontrivial symmetry algebra exists (with $N=1$)  and the ODE can be integrated in terms of elliptic or elementary functions. If $q(x)$ is itself a Painlev\' e transcendent, then so is $y$ and we have  $dim L=0$.
\end{enumerate}

Let us now look at the $6$ irreducible Painlev\' e transcendents in detail and calculate their Lie point symmetry algebras. We already know that for generic values of their parameters their Lie point symmetry algebras are $L=\{0\}$ in all cases. Applying the standard symmetry algebra algorithm we find that for $P_I$, $P_{II}$ and $P_{IV}$  we have $L=\{0\}$ for all values of the parameters in $P_{II}$ and $P_{IV}$  ($P_I$ has no parameters). For $P_{III}$, $P_V$  and $P_{VI}$ the situation is different. Let us consider the three cases separately.
\begin{enumerate}
\item   $\boldsymbol{P_{III}}$. The ODE in its standard form is:
\bea \label{2.9}
y''=\frac 1 y (y')^2
- \frac 1 x y' + \frac 1 x (\alpha y^2 +\beta) + \gamma y^3 + \frac{\delta} y
\eea
Nontrivial Lie algebras are obtained in the following cases
\begin{enumerate}
\item $\alpha=\beta=\gamma=\delta=0$. We obtain $dim L=8$ realized by the vector fields
\bea \label{2.10}
\hat Y_1=y \partial_y, \quad \hat Y_3=y \ln y \partial_y, \quad \hat Y_5=y \ln x \partial_y, \quad \hat Y_7=x \ln x \ln y \partial_x + y \ln^2 y \partial_y, \\ \nonumber
\hat Y_2=x \partial_y, \quad \hat Y_4=x \ln x \partial_x, \quad \hat Y_6=x \ln y \partial_x, \quad \hat Y_8=y \ln x \ln y \partial_y + x \ln^2 x \partial_x
\eea
By a change of variables
\bea \label{2.11}
(x,y) \rightarrow (t,z), \; t=\ln x, \; z=\ln y,\;x=e^t, \; y=e^z,
\eea
we transform the algebra $sl(3)$ into a more familiar form, namely
\bea \label{2.12}
\hat Y_1= \partial_z, \quad \hat Y_3=z  \partial_z, \quad \hat Y_5=t \partial_z, \quad \hat Y_7=z(t  \partial_t + z  \partial_z), \\ \nonumber
\hat Y_2= \partial_t, \quad \hat Y_4=t  \partial_t, \quad \hat Y_6=z  \partial_t, \quad \hat Y_8=t( t \partial_t + z  \partial_z).
\eea
This is the Lie algebra of the group of projective transformations of the Euclidean plane $E_2$ (over $\mathbb C$ or $\mathbb R$).
Eq. \eqref{2.9} is transformed into
\bea \label{2.13}
z_{tt}= e^t(\alpha e^z + \beta e^{-z}) + e^{2t} ( \gamma e^{2z} + \delta e^{-2z}),
\eea
so for $\alpha=\beta=\gamma=\delta=0$ we obtain 
\bea \label{2.14}
z_{tt}=0, \quad z(t)= A t + B,
\eea
with symmetry algebra $sl(3)$ realized as \eqref{2.12}. Transforming the solution \eqref{2.14} back to the original variables, we obtain
\bea \label{2.15}
y=c_1 \, x^{c_2}, \quad ( c_1=e^B, \; c_2=A).
\eea
The other cases with a nontrivial symmetry algebra are
\item $\alpha \ne 0, \; \beta=\gamma=\delta=0, \; dim L=2$
$$ \hat A = \hat Y_1-\hat Y_2=\partial_z-\partial_t, \quad \hat B = \hat Y_4-\hat Y_5-2 \hat Y_1=-t(\partial_z - \partial_t)-2 \partial_z$$
\item $\beta \ne 0, \; \alpha=\gamma=\delta=0, \; dim L=2$
$$ \hat A = \hat Y_1+\hat Y_2=\partial_z+\partial_t, \quad \hat B = \hat Y_4+\hat Y_5+2 \hat Y_1=t(\partial_z + \partial_t)+2 \partial_z$$
 \item $\gamma \ne 0, \; \beta=\alpha=\delta=0, \; dim L=2$
$$ \hat A = \hat Y_1-\hat Y_2=\partial_z-\partial_t, \quad \hat B = \hat Y_4-\hat Y_5- \hat Y_1=-t(\partial_z - \partial_t)- \partial_z$$
\item $\delta \ne 0, \; \beta=\gamma=\alpha=0, \; dim L=2$
$$ \hat A = \hat Y_1+\hat Y_2=\partial_z+\partial_t, \quad \hat B = \hat Y_4+\hat Y_5+ \hat Y_1=t(\partial_z + \partial_t)+ \partial_z$$
\item $\beta=\delta=0,\,\alpha \ne 0, \, \gamma \ne 0, \; dim L=1, \; \hat A = \hat Y_1 - \hat Y_2 = \partial_z - \partial_t,$
\item $\alpha=\gamma=0,\,\beta \ne 0, \, \delta \ne 0, \; dim L=1, \; \hat A = \hat Y_1 + \hat Y_2 = \partial_z + \partial_t.$
\end{enumerate}
The two dimensional algebra in all the above cases satisfies $[\hat A, \hat B]=\hat A$.

To integrate the ODE corresponding to cases (b),$\cdots$,(g) above we need a further transformation $(t,z) \rightarrow (s,u)$ such that 
\bea \label{2.16}
\hat A = \partial_u, \qquad \hat B= s \partial_s + u \partial_u,
\eea
to take \eqref{2.13} with the parameters appropriately specified in each case to its {\it standard form}. This is $\ddot u = F(s, \dot u)$ for $dim L=1$, $\ddot u = \frac 1 s F(\dot u)$ for $dim L=2$, where for \eqref{2.9} $F$ is some specific function. Alternatively, once the values of $\alpha,\, \beta,\,\gamma,\,\delta$ have been specified, one can simply ask Maple to integrate \eqref{2.9}. Indeed the Lie group procedure has been built into the routines for analytically solving ODEs. The Maple procedure is based on the articles \cite{ct1,ct2}.

In all cases the solutions are elementary and quite explicit. For example, for case (b) the two dimensional Lie algebra leads to the solution
\bea \label{2.17}
y(x) ={\frac {1}{2{{c_1}}^{2}\alpha\,x} \left(     \tan^2 \left( {\frac {\ln  \left( x \right) +c_2}{2 c_1}} \right)  +1 \right) }
\eea
\item $\boldsymbol {P_V}$. The ODE is 
\bea \label{2.18}
y''= \Big ( \frac 1 {2y} + \frac 1 {y-1} \Big ) (y')^2 -\frac 1 x y' + \frac{(y-1)^2}{x^2}\Big (\alpha y + \frac {\beta} y \Big) + \gamma \frac y x + \delta \frac{y(y+1)}{y-1}
\eea
Nontrivial Lie point symmetry algebras are obtained in the following cases
\begin{enumerate}
\item $\alpha=\beta=\gamma=\delta=0$; $dim L=8$

The algebra is realized by the vector fields
\bea \nonumber
&&\hat Y_1= \sqrt{y}(y-1)\partial_y, \quad \hat Y_3=\sqrt{y}(y-1) \mbox{arctanh}(\sqrt{y}) \partial_y,  \\  \label{2.19}
&&\hat Y_2=x \partial_x, \quad \hat Y_4=x \ln x  \partial_x, \quad \hat Y_5=\ln x \sqrt{y}(y-1) \partial_y,  \\ \nonumber &&  \hat Y_6=x  \mbox{arctanh} (\sqrt{y})  \partial_x, 
\\ \nonumber &&\hat Y_7=-\frac 1 2 x \ln x \mbox{arctanh} (\sqrt{y}) \partial_x +\sqrt{y}(y-1) \mbox{arctanh}^2(\sqrt{y}) \partial_y, 
\\ \nonumber &&\hat Y_8=- \frac 1 2 x \ln^2 x \partial_x + (y-1)\sqrt{y} \,\ln x \, \mbox{arctanh}(\sqrt{y}) \partial_y.
\eea
This realization of $sl(3)$ is taken into the standard form \eqref{2.12} by the transformation
\bea \label{2.20}
(x,y) \rightarrow (t,z), \, t=\ln x, \, z=2 \mbox{arctanh}(\sqrt{y}), \, x=e^t, \, y=\tanh^2 \frac z 2.
\eea
The transformation \eqref{2.20} take \eqref{2.18} into $z_{tt}=0$ so we obtain the solution $z=At+B$ which in this case means 
\bea \label{2.20a} 
y(x) = \tanh^2 \Big (\frac A 2 \ln x + B \Big)= \Big ( \frac{x^{2A} e^{2B} -1}{x^{2A} e^{2B} +1} \Big )^2
\eea
 (we see the movable double pole at $x=-e^{-\frac B A}$).

\noindent The other special case with a nontrivial symmetry algebra is
\item  $\gamma=\delta=0,\,(\alpha, \beta) \ne (0, )), \; dim L=1, \; \hat Y =  x\partial_x$ 

This again leads to an elementary solution.
\end{enumerate}
\item $\boldsymbol {P_{VI}}$. The ODE is 
\bea \label{2.21}
\ddot y  &=&\frac 1 2\, \left( 
\frac 1 y + \frac 1{y-1} + \frac 1 {y-x} \right) 
{\dot y}^{2} - \left ( \frac 1 x + \frac 1 {x-1} + \frac 1 {y-x} \right ) \dot y \\ \nonumber 
&+& \frac{y(y-1)(y-x)}{x^2(x-1)^2} \left [ \alpha + \beta \frac x {y^2} + \gamma  \frac {x-1}{(y-1)^2} + \delta \frac{x(x-1)}{(y-x)^2} \right ]
\eea
The determining equations for the symmetry algebra are quite complicated in this case. So we used the Rif (Reduced involutive form) simplification algorithm contained in the Maple symbolic manipulation  program \cite{R,RWB} and we obtain that only for the values of the parameters 
$$ \alpha=\beta=\gamma=0, \quad \delta=\frac 1 2$$
the determining equations for the symmetries are solvable and tell us that the dimension of the symmetry algebra is $dim L=8$.

The linearizing transformation is known \cite{Pic,11,12,Mazz,CD} and involves elliptic integrals and solutions of the Heun equation.
\end{enumerate}


\section{Higher order ODEs}
The theory and specially the classification of ODEs of order $n \ge 3$ with the Painlev\' e property is much less developped than for $n=2$.

Early work is due to Bureau \cite{18,19,20}, mainly on the case $n=3$. Much of the subsequent work was concentrated on equations in the {\it Bureau polynomial class} where the function $R$ in \eqref{1.1} is assumed to be polynomial, rather than a more general rational function. Classical results are due to Chazy \cite{21,22} and Garnier \cite{23}. More recent is a very useful series of articles by Cosgrove \cite{24,25,26,27,28,29,30,31}. For ODEs that go beyond the polynomial class see e.g. Mugan et. al. \cite{32,33,34} and Kudryashov \cite{35,36}. 

Here we shall just concentrate on some examples that are candidates for being new higher order Painlev\' e transcendents. We do not consider those that have already been integrated, or that have obvious symmetries (like autonomous equations with their $\hat X=\partial_x$ translational symmetry). In all considered
cases it turns out that the symmetry algebra contains only translations
in $x$ and dilations in $x$ and $y$, i.e. elements of the form  $\hat X =\partial_x$ and $\hat Y= A x \partial_x +B y \partial_y$ Where $A$ can be restricted to $A=1$ or $A=0$ and $B$ is
a number to be specified in each case. The corresponding group
transformations are
\bea \label{3.0}
 \tilde x = e^{\lambda\, A}x,  \quad \tilde y = e^{\lambda\,B}y, 
\eea
where $\lambda$ is a group parameter.

{\bf Third order ODEs}.  To our knowledge all third order ODEs in the polynomial class that have the Painlev\' e property have already been integrated in terms of lower order transcendents, so we shall concentrate on nonpolynomial ones given by Mugan and Jrad in \cite{33}. They considered ODEs of the form 
\bea \label{3.1}
y'''=c_1 \frac{y' y''} y + c_2 \frac {(y')^3}{y^2} + F(y, y', y'';x),
\eea
where $c_1$ and $c_2$ are constants with $(c_1,c_2)\ne(0,0)$. They found many such equations and integrated most of them. Others are autonomous, so their order can be decreased by at least one. Notice that for any values of the
constants $c_1$ and $c_2$ the first three terms in \eqref{3.1} transform  identically under
dilations, i.e. they are multiplied by $e^{(A-B)\, \lambda}$. Hence
for F=0 the symmetry algebra is three-dimensional: $\{\hat X_1 =\partial_x, \, \hat X_2 =x\partial_x,\, \hat X_3
=y\partial_y \}$.

Let us look at some of the remaining ones. The equations are numbered as in \cite{33}; the $k_i$ are constants (the boldface
equation numbers before the equations are from the original article \cite{33}).

\noindent {\bf Eq. 2.67}
\bea \label{3.2}
y'''&=&4\,\frac {
 y' y'' }{y }-3\,\frac { (y') ^3}{  y^2}+ y^2 y' +\frac { \left( {\it k_3}\,x+{\it k_2} \right) y'}{y}+\frac {{\it k_1}
\,y'}{  y^2}\\ \nonumber &&-\frac 3 4\,{\frac {{\it k_3}\, \left( {\it k_3}\,x+{
\it k_2} \right) y}{{\it k_1}}}-{\it k_3}
\eea
For $(k_1,\,k_3) \ne (0,0)$ no symmetry.

\noindent For $k_1 \ne 0$ and $k_3=0$ $\hat X=\partial_x$.

\noindent For
$k_1 \ne 0$ and $k_2=k_3 =0$ we have the symmetry algebra $\{\hat X_1 =\partial_x,  \hat X_2 = y\partial_y-x\partial_x\}$.

\noindent {\bf Eq. 2.106}
\bea \label{3.3}
 y''' = \frac{y' y''} y -2 y y'' + y^2 y' +y^4 +(k_2x+k_3)y^2 +k_2\Big (\frac{ 2y'} y +y \Big )
\eea
For general $k_2,\,k_3$: no symmetry.

\noindent For $k_2=0,\,k_3$ arbitrary: $\hat X= \partial_x$.

\noindent For $k_2=k_3=0$ we have the symmetry algebra $\{\hat X_1=\partial_x, \; \hat X_2=x \partial_x - y \partial_y\}$.

\noindent {\bf Eq. 4.6}
\bea \label{3.4}
y''' =3 \frac{y'y''}y - \frac{16 (y')^3}{9 y^2} +(k_1x+k_2) y' + k_2y=0
\eea
For any $k_1$ and $k_2$ we have the symmetry $\hat X= y \partial_y$. The transformation $y=e^u$ takes $\hat X$ into $\hat X=\partial_u$. Putting $u_x=3w$ we reduce \eqref{3.4} to the Painlev\' e transcendent $P_{II}$ for $k_1 \ne 0$. For $k_1=1$ the algebra is $\{\hat X_1=\partial_x, \, \hat X_2= y \partial_y \}$ and the ODE reduces  to that for elliptic functions.

\noindent {\bf Eq. 4.14}
\bea \label{3.5}
y'''=\frac{y' y''} y -24 y^2 +k_1y+\Big(\frac{k_1^2}{12}x + k_2 \Big ) \frac{y'} y - \frac{k_1^2}{12}
\eea
For $k_1,\,k_2$ general, no symmetries.

\noindent For $k_1=0$, $\hat X_1=\partial_x$

\noindent For $k_1=k_2=0$ we have the symmetry algebra  $\{\hat X_1=\partial_x,\;\hat X_2=x\partial_x-3 y \partial_y\}$.

We see that \eqref{3.2}, \eqref{3.3} and \eqref{3.5} survive as candidates for new Painlev\'e transcendents (for generic values of the parameters $k_i$), whereas \eqref{3.4} is integrated in terms of lower order transcendents.

{\bf ODEs of order 4 not in polynomial class}  We take 3 such ODEs from the article \cite{35} by Kudryashov.

\noindent {\bf Eq. 3.14}
\bea \label{3.6}
&&y'''' - 3 \frac{y'y''}y - \frac72\frac{(y'')^2}y+\frac{17}2 \frac{(y')^2y''}{y^2}-\frac{27}{8}\frac{(y')^4}{y^3}+\Big ( \beta -\frac{5\delta}{y}\Big ) y''\\ \nonumber &&-\frac 12\Big(\frac{\beta}y-\frac{15 \delta}{y^3} \Big ) (y')^2 + 2 \gamma y^2 - 2 \alpha x y +\frac{\beta\delta}y -\frac{3 \delta^2}{2 y^3}=0
\eea
For $\alpha$, $\beta$, $\gamma$, $\delta$ generic there are no symmetries.

\noindent For $\alpha \ne 0$, $\beta$ arbitrary, $\gamma=\delta=0$ the algebra is $\hat X  =y \partial_y$.

\noindent For $\alpha=0$. $\beta \ne 0$, $\gamma \ne 0$, $\delta=0$ we have the symmetry algebra $\{\hat X_1 =\partial_x,  \hat X_2 = x \partial_x - 4 y \partial_y\}$.

\noindent For $\alpha=\beta =\gamma=0$, $\delta \ne 0$ we have the symmetry algebra $\{\hat X_1 =\partial_x,  \hat X_2 = x \partial_x +y \partial_y\}$.

\noindent For
$\alpha=\beta=\gamma=\delta=0$ we have the symmetry algebra $\{\hat X_1 =\partial_x,  \hat X_2 = x \partial_x, \, \hat X_3=y \partial_y\}$.

\noindent {\bf Eq. 3.28}
\bea \label{3.7}
&&y''''-4\frac{y' y'''}y-3\frac{(y'')^2}y+\frac{21}2 \frac{(y')^2y''}{y^2}-\frac 92\frac{(y')^4}{y^3}-\Big (2\alpha x +\frac{5\delta}{y^2} \Big)y''\\ \nonumber &&+2 \Big ( \frac{\alpha x}y+\frac{5 \delta}{y^3} \Big ) (y')^2+\beta y^2 -2 \alpha y' +\gamma - 4 \frac{\alpha \delta x}y - 2 \frac{\delta^2}{y^3}=0
\eea
For $\alpha$, $\beta$, $\gamma$, $\delta$ generic there are no symmetries.

\noindent For $\alpha \ne 0$, $\beta=\gamma=\delta=0$ the algebra is $\hat X= y \partial_y$.

\noindent For $\alpha=0$. $\beta \ne 0$, $\gamma =\delta=0$ we have the symmetry algebra $\{\hat X_1 =\partial_x,  \hat X_2 = x \partial_x - 4 y \partial_y\}$.

\noindent For $\alpha=\beta =0$, $\gamma \ne 0$, $\delta = 0$ we have the symmetry algebra $\{\hat X_1 =\partial_x,  \hat X_2 = x \partial_x +y \partial_y\}$.

\noindent For
$\alpha=\beta=\gamma=\delta=0$ we have the symmetry algebra $\{\hat X_1 =\partial_x,  \hat X_2 = x \partial_x, \, \hat X_3=y \partial_y\}$.

\noindent {\bf Eq. 4.6}
\bea \label{3.8}
&&y''''-2\frac{y' y'''}y -5y^2y''-\frac 52 y (y')^2-\frac 32 \frac{(y'')^2}y+2\frac{(y')^2y''}{y^2}-2\frac{\alpha x y''}y\\ \nonumber &&+2\frac{\alpha x (y')^2}{y^2}-2 \frac{\alpha y'}y+\frac 52 y^5 -\beta^2y^3+2 \alpha x y^2 +\gamma y -\frac 12 \frac{\alpha^2 x^2}y=0
\eea
For $\alpha \ne 0$ no symmetry.

\noindent For $\alpha=0$ $\hat X=\partial_x$.

\noindent For $\alpha=0$, $\beta$ and $\gamma$ arbitrary, the algebra is $\hat X= \partial_x$.

\noindent For
$\alpha=\gamma =0$, $\beta$ arbitrary  we have the symmetry algebra $\{\hat X_1 =\partial_x,  \hat X_2 = x\partial_x-y\partial_y\}$.


{\bf ODEs of order 4 and 5 in the polynomial class}. We use the Cosgrove classification presented in \cite{30} and \cite{31} and calculate the Lie point symmetries of the ODEs that he identifies as defining new transcendents. Specifically these are equations 
\bea \label{3.9} 
\boldsymbol{ F_V}:\quad y''''=20 y y'' +10 (y')^2 -40 y^3 +\alpha y +k x +\beta
\eea
\bea \label{3.10}
\boldsymbol{F_{VI}}: \quad y''''= 18 y y'' +9 (y')^2-24 y^3 +\alpha y^2 +\frac 19 \alpha^2 y +k x +\beta
\eea
\bea \label{3.11}
\boldsymbol{F_{XVII}}: \quad  y''''=10 y^2 y'' + 10 y (y')^2 - 6 y^5 -10 \delta(y''-2y^3)+(\lambda x + \mu )y + \sigma
\eea
\bea \label{3.12}
\boldsymbol{F_{XVIII}}: \quad y''''=-5y' y'' +5 y^2 y'' +5 y (y')^2 - y^5 +(\lambda x + \alpha)y + \gamma
\eea
\bea \label{3.13}
\boldsymbol{\mbox{{\bf Fif}}_{IV}}: \quad && y'''''=18 y y''' +36 y' y'' -72 y^2 y' + 3 \lambda y''+\frac 12 \lambda x (5 y''' -36 y y')\\ \nonumber && + \frac 1 x \{ y'''' -18 y y''-9(y')^2+24 y^3 -3 \lambda y' + k \} -\frac 12 \lambda^2 x(2 x y' +y)
\eea
All of them depend on 2 or 3 parameters and for general values of the parameters none of them has a nontrivial symmetry algebra.

For specific values of the parameters, the situation is as follows:

\noindent Eq. \eqref{3.9}: $\alpha=\beta= k=0, \qquad \hat A= \partial_x, \; \hat B= x\partial_x -2 y \partial_y$

$k=0,\, \alpha, \, \beta$ arbitrary, $\hat A=\partial_x$

\noindent Eq. \eqref{3.10}: $\alpha=\beta= k=0, \qquad \hat A= \partial_x, \; \hat B= x\partial_x -2 y \partial_y$

$k=0,\, \alpha, \, \beta$ arbitrary, $\hat A=\partial_x$

\noindent Eq. \eqref{3.11}: $\lambda=\mu= \delta=\sigma=0, \qquad \hat A= \partial_x, \; \hat B= x\partial_x - y \partial_y$

$\lambda=0,\, \delta, \, \mu, \, \sigma$ arbitrary, $\hat A=\partial_x$

\noindent Eq. \eqref{3.12}: $\alpha=\lambda= \gamma=0, \qquad \hat A= \partial_x, \; \hat B= x\partial_x - y \partial_y$

$\lambda=0,\, \alpha, \, \gamma$ arbitrary, $\hat A=\partial_x$

\noindent Eq. \eqref{3.13}: $\lambda= k=0, \qquad  \hat A= x\partial_x -2 y \partial_y$

The situation is very similar to that of $P_{III}$ and $P_{V}$ for second order ODEs. In general these ODEs define new transcendents. For special values of the parameters lie group theory makes it possible to reduce to lower order.

\section{Conclusions}
From the analysis of ODEs passing the Painlev\'e test we can draw the following conclusions valid for ODEs of any order.
\begin{enumerate}
\item If an ODE that is a candidate for having the Painlev\'e property has a symmetry algebra of dimension $dimL \ge 1$ it can by a change of variables be reduced to an ODE of lower order. The reduction $(x, y) \rightarrow (t, u)$ does not necessarily preserve the Painlev\'e property. However, if the reduced equation can be solved for $u(t)$, then the inverse transformation $u(t) \rightarrow y(x)$ will produce solutions of the original equations and the Painlev\'e property is restored. In any case, if we are looking for new Painlev\'e transcendents, not expressible in terms of lower order ones, only equations with $L = \{0\}$ need to be considered. 
\item Autonomous equations $ y^{(n)} (x) = R(y, y’ , \cdots , y^{(n-1)})$ always have at least a one dimensional symmetry algebra, $L = \{\partial_x\}$. Hence autonomous equations can never produce new transcendents.
\item If a non-autonomous equation depends on parameters and in general has $L = \{0\}$, then group analysis can pick values of the parameters for which we have $dimL \ge 1$. For those values the order of the ODE can be reduced and sometimes solutions can be obtained in terms of known functions.
\end{enumerate}

A complete classification of ODEs of the form (1) that have the Painlev\'e property exists only for $n=1$, and $n=2$. For $n>2$ the classifications are very incomplete and most authors concentrated on the polynomial class of equations [7,8,9, 16-23]. That notwithstanding, higher order ODEs passing the Painlev\'e test keep appearing in physical applications, in particular as superintegrable potentials in quantum mechanics. These are Hamiltonian systems that allow more integrals of motion than they have degrees of freedom. In quantum mechanics it is usually assumed that the integrals are polynomials of order $N$ in the momenta, i.e. differential operators of order $N$. For $N\ge 3$ {\it exotic} potentials start appearing. They do not satisfy any linear differential equation, only nonlinear ones. For $N= 3,4,5$ it turned out that these equations always passed the Painlev\'e test. For $N=3$ and $N=4$ the equations were integrated in terms of known  (second order) Painlev\'e transcendents.  For $N\ge5$ new {\it irreducible} higher order transcendents start appearing. It has been conjectured that superintegrable exotic potentials with integrals of any order will have the Painlev\'e property [4,25,38]. 

    It is doubtful that the complete classification of Painlev\'e type equations can be pushed much further.  This makes it all the more important to have tools for obtaining special solution and for lowering the order of the nonlinear equations. 

    In this article we have restricted ourselves to Lie point symmetries. We have shown that they can help to identify Painlev\'e type equations that lead to lower order transcendents, to identify cases when the nonlinear equation can be linearized and to find particular “invariant” solutions.

   More general symmetry transformations, for instance Lie--B\"acklund ones, should give further results. In particular they give many families of classical solutions of the original Painlev\'e equations $P_{II}, \ldots, P_{VI}$ (for special values of the parameters) \cite{gromak}.

After completing this manuscript (and publishing its preprint as arXiv: 1712.09811v1) we became aware of a preprint by Contatto and Dunajski \cite{CD}. They address and solve a different problem, namely which of the second order ODEs having the Painlev\'e property are metrisable (i.e. their integral curves are geodesics of a pseudo Riemannian metric on some surface). In the Appendix we present the relation between their and our results.

\paragraph{Acknowledgments.}   DL has been supported by  INFN IS-CSN4 {\it Mathematical Methods of Nonlinear
Physics}. The research of PW was
partially supported by a discovery research grant from NSERC of Canada. DL thanks the CRM for the support during his stay in Montr\'eal where this research has been carried out. We thank  M. Dunajski for calling the preprint \cite{CD} to our attention and for very helpful (electronic) discussions. We thank E. Cheb--Terrab  for helping us in running the Rif program.

\appendix
\section*{Appendix}

In this Appendix we present the interesting relation between the results,  contained in \cite{CD}, from now on indicated as {\it their results} and our results.

Their results are summed up in Theorem 1.3 and in five points ({\it bullets}) in the conclusions of \cite{CD}. Let us compare them with the results presented here.

The Theorem 3.1 of \cite{CD} concerns the 6 {\it irreducible} Painlev\'e transcendents and states that the only metrisable ones are $P_{III}$ for $\alpha=\gamma=0$, or $\beta=\delta=0$, $P_V$ for $\gamma=\delta=0$ and $P_{VI}$ for $\alpha=\beta=\gamma=0$ and $\delta=\frac 1 2$.  Moreover the projective structure is flat for $P_{III}$ and $P_V$ with $\alpha=\beta=\gamma=\delta=0$ and for $P_{VI}$ with $\alpha=\beta=\gamma=0, \delta=\frac 1 2$.

Our result is $dim L=8$ for $P_{VI}$ iff $\alpha=\beta=\gamma=0, \delta= \frac 1 2$ and for $P_{III}$ and $P_V$ with $\alpha=\beta=\gamma=\delta=0$. We obtain $dim L=1$ or $2$ in the other metrisable cases.

The results contained in \cite{CD} for all 50 equations are related to ours as follows (we follow the list contained in \cite{CD} in Section 4). 
\begin{enumerate}
\item Metrisable with $4 > m > 1$. In our list they all have $dim L=1$ or $2$ and are in our Class 1 (autonomous equations).
\item Metrisable with $m=4$. In our list in Class 1 but with $dim L=3$.
\item Not metrisable, with a degenerate solution. We have  $dim L=0$ in all cases. We mention that $I_{14}$ is reducible to a Riccati equation and is in our Class 2. The other cases, $I_{20}$ and $I_{34}$ are in our Class 3 and are equivalent to the transcendent $P_{II}$.
\item Not metrisable. Class 3 and 4 in our list. They all involve unknown functions. In general have $dim L=0$ but may be nontrivial for some specific functions.
\item Metrisable and projectively flat. In our list: $dim L=8$ ($L \sim sl(3,\mathbb R)$). All in our Class 1 except $I_6$ which is in Class 3.
\end{enumerate} 

\end{document}